\documentclass{aa}
\usepackage{graphicx}
\usepackage{txfonts}
\usepackage{booktabs}
\usepackage{siunitx}
\usepackage{hyperref}
\hypersetup{
    colorlinks=true,
    linkcolor=blue,
    filecolor=magenta,      
    urlcolor=blue,
    citecolor=blue
} %to customize the colors of references and hyperlinks in general

\bibpunct{(}{)}{;}{a}{}{,} % to follow the A&A style
\usepackage{longtable}
\usepackage{threeparttable}
\usepackage{amsmath}
\usepackage{upgreek,textcomp}
\usepackage{multicol}
\usepackage{placeins}

\begin{document}

\title{Multi-scale VLBI observations of the candidate host galaxy of GRB~200716C}

\subtitle{}

\author{S.~Giarratana \inst{1,2}\fnmsep\thanks{E-mail: {stefano.giarratana2@unibo.it}} \and M.~Giroletti \inst{2} \and C.~Spingola \inst{2} \and G.~Migliori \inst{2} \and S.~Belladitta \inst{3,4} \and M.~Pedani \inst{5}}

\institute{Dipartimento di Fisica e Astronomia, Universit\`{a} degli Studi di Bologna, Via Gobetti 93/2, I$-$40129 Bologna, Italy
\and
INAF $-$ Istituto di Radioastronomia, Via Gobetti 101, I$-$40129, Bologna, Italy
\and
INAF $-$ Osservatorio Astronomico di Brera, Via Brera, 28, 20121 Milano, Italy
\and
DiSAT, Universit\`a degli Studi dell'Insubria, Via Valleggio 11, 22100 Como, Italy
\and
Fundación Galileo Galilei - INAF, Rambla José Ana Fernández Pérez, 7 38712 Breña Baja, TF, Spain 
}
\date{Received \dots; accepted \dots}

\abstract
% context heading (optional)
{
We present the discovery and the subsequent follow up of radio emission from SDSS J130402.36$+$293840.6 (J1304+2938), the candidate host galaxy of the gamma-ray burst (GRB) GRB~200716C. The galaxy is detected in the RACS (0.89\,GHz), the NVSS, the Apertif imaging survey, and the FIRST (1.4\,GHz), the VLASS (3\,GHz), and in public LOFAR (130-170\,MHz), WISE (3.4-22\,$\upmu$m), and SDSS ($z, i, r, g, u$ filters) data. The luminosity inferred at 1.4\,GHz is (5.1$\pm0.2 )\times 10^{30}$\,erg\,s$^{-1}$\,Hz$^{-1}$. To characterise the emission and distinguish between different components within the galaxy, we performed dedicated, high-sensitivity and high-resolution observations with the European VLBI Network (EVN) + \textit{e}-MERLIN at 1.6 and 5\,GHz. We did not detect any emission from a compact core, suggesting that the presence of a radio-loud active galactic nucleus (AGN) is unlikely, and therefore we ascribe the emission observed in the public surveys to star-forming regions within the galaxy. We confirm and refine the redshift estimate, $z = 0.341 \pm 0.004$, with a dedicated Telescopio Nazionale Galileo (TNG) spectroscopic observation. Finally, we compiled a list of all the known hosts of GRB afterglows detected in radio and computed the corresponding radio luminosity: if GRB~200716C belongs to J1304+2938, this is the third most radio-luminous host of a GRB, implying one of the highest star-formation rates (SFRs) currently known, namely SFR $\sim$ 324$\pm$61\,M$_{\odot}$\,yr$^{-1}$. On the other hand, through the analysis of the prompt emission light curve, recent works suggest that GRB~200716C might be a short-duration GRB located beyond J1304$+$2938 and gravitationally lensed by an intermediate-mass black hole (IMBH) hosted by the galaxy. Neither the public data nor our Very Long Baseline Interferometry (VLBI) observations can confirm or rule out the presence of an IMBH acting as a (milli-)lens hosted by the galaxy, a scenario still compatible with the set of radio observations presented in this work.}


\titlerunning{VLBI observations of J1304+2938}
\maketitle

\section{Introduction}
Gamma-ray bursts (GRBs) are cosmological explosions whose emission spans the whole electromagnetic spectrum, from soft $\gamma$-rays down to X-rays, optical/near-infrared(NIR), and radio (see e.g. \citealt{Piran2004}). 
According to the T$_{90}$ duration of their short-lived prompt emission, they are classified as short-duration (T$_{90}<$ 2\,s) and long-duration GRBs (T$_{90}\ge$ 2\,s; \citealt{Kouveliotou1993}). This (apparently) arbitrary and crude separation has a deep connection with the progenitor's nature of the burst: while short-duration GRBs flag the merger of two neutron stars or a neutron star and a black hole, as confirmed by the outstanding detection of the first multi-messenger event GW~170817/GRB~170817A \citep{Abbott2017}, long-duration GRBs are produced in the catastrophic explosion of massive single stars, as confirmed by many long-duration GRBs associated with supernovae \citep[SNe; see, e.g.][]{Galama1998, Hjorth2003, Stanek2003}. The different nature of the progenitors is further corroborated by the dichotomy between the hosts of short- and long-duration GRBs: while {all} morphological types of galaxies can harbour a short-duration GRB \citep{Berger2009, Fong2013, Berger2014}, in agreement with the fact that binaries are expected to be widespread, long-duration GRBs are found predominantly in highly star-forming regions (\citealt{Berger2014,Klose2019}, and references therein), as expected from a parent population of young massive stars.

Studying GRB host galaxies is therefore crucial for directly investigating the nature of the progenitor, its formation channel, and the circumburst medium. In particular, radio and submillimeter observations can be useful for determining the level of obscured star formation and the overall properties of highly star-forming galaxies at high redshifts, such as metallicity and star formation rate \citep[SFR;][]{Berger2001}, or the interaction between the host galaxy and the surrounding intergalactic medium \citep{Stanway2015, Michalowski2015}. The first study of the radio properties of GRB host galaxies was performed by \citet{Berger2003}: the authors studied 20 sources and found that the SFR inferred from the radio measurements exceeds the values determined from the optical by an order of magnitude, suggesting significant dust obscuration.
Conversely, \citet{Stanway2010} observed a sample of five galaxies and found a radio-derived SFR $<$ 15\,M$_{\odot}$\,yr$^{-1}$, in agreement with the values inferred from optical estimators, suggesting little dust obscuration. Other studies tackled this problem \citep{Berger2001, Berger2003, Michalowski2012, Hatsukade2012, Perley2013, Stanway2014, Stanway2015, Perley2015, Michalowski2015, Greiner2016} and, although they generally agree with the hypothesis of little dust obscuration, a conclusive result is still missing due to the dearth of detected sources: among the approximately 87 host galaxies that have been observed in the radio, only 20 have a confirmed detection, corresponding to a $\sim 23$\% detection rate. As a consequence, outstanding questions remain unanswered, such as whether or not long GRBs are unbiased tracers of the cosmic star formation history, or whether or not they provide clues as to a particular formation channel of young massive stars \citep{Berger2001, Ghirlanda2022}, .

A complementary approach is based on the use of ongoing radio sky surveys provided by the Square Kilometre Array (SKA) precursors and pathfinders, such as the Rapid Australian SKA Pathfinder Continuum Survey \citep[RACS;][]{McConnell2020}, the Very Large Array Sky Survey \citep[VLASS;][]{Lacy2020}, and the LOw-Frequency ARray (LOFAR) Two-metre Sky Survey \citep[LoTSS;][]{Shimwell2017}. The rms noise levels of these surveys are seldom deep enough to reveal faint radio emission from GRB hosts; however, they provide a handy resource with which to carry out a systematic search, which is ideal for singling out the most extreme objects for subsequent follow up with dedicated observations. In this paper, we follow this approach and present a detailed radio study of the candidate host galaxy of GRB~200716C based on public survey data and new, dedicated, deep and high angular resolution radio observations.

%Because of their high luminosities (up to 10$^{53-54}$\,erg\,s$^{-1}$ \citep{Piran2004, Kumar2015}, GRBs can be detected up to the highest refshifts: the farthest GRB current known is GRB 090429B at a photometric redshift of $z = 9.4$ \citep{Cucchiara2011}. Therefore, they can be used as probes of the early Universe \citep{Fryer2021}.
%As GRBs could be cosmologically distant events, some of them might be gravitationally lensed (e.g., \citealt{Paynter2021} and references therein). Because of the strong lensing effect, photons coming from a distant source travel different geometric paths as they approach the foreground lensing object and form multiple magnified images of the same background source \citep{Congdon2018}.
%As a consequence, we observe variations in the lensed images with a time delay, which depends on the gravitational potential of the lens. In the case of GRBs, if gravitationally lensed, we expect to measure a bright $\gamma$-ray pulse followed by a dimmer duplicate.
%
%To date only a few GRBs have been suggested as candidate lensed events, namely GRB~950830 \citep{Paynter2021}, GRB~210812A \citep{Veres2021}, GRB~081126A and GRB~090717A \citep{Lin2021}, based on the analysis of their light curves. Among these, 
GRB~200716C triggered the \emph{Fermi} Gamma-ray Burst Monitor (GBM) at 22:57:41 UT on 2020 July 16, %mjd = 59046
which classified it as a long-duration GRB \citep{Fermi2020, Veres2020}. The prompt emission was subsequently detected by \emph{Swift} Burst Alert Telescope (BAT) and X-Ray Telescope (XRT; \citealt{Ukwatta2020}), \emph{AGILE} Mini-CALorimeter \citep{Ursi2020}, \emph{CALET} Gamma-Ray Burst Monitor \citep{Torii2020}, Insight-HXMT/HE \citep{Xue2020}, and Konus/Wind \citep{Frederiks2020}. \citet{D'Avanzo2020} detected an extended source in the Sloan Digital Sky Survey (SDSS) within $\sim1$\,arcsec from the location of the optical afterglow of GRB~200716C, and they estimated a photometric redshift of $z=0.348\pm0.053$ for SDSS J130402.36$+$293840.6 (J1304$+$2938 hereafter). Other optical detections of this galaxy were subsequently reported \citep{Kumar2020, Pozanenko2020, Kann2020}.
%=== PORTARE NELLA DISCUSSIONE ===
%Based on the photometric redshfit, \citet{Frederiks2020} found that GRB~200716C is a clear outlier in the Amati relation \citep{Amati2002} for 138 Konus/Wind\footnote{Konus is a GRB monitor launched on the GGS-Wind spacecraft in November 1994.} long-duration GRBs.  Offsets of about $\sim2$\,dex in energy from the main Amati correlation typically indicate short-duration GRBs \citep[e.g.,][]{Willingale2017}.
%Because of this and the presence of two pulses separated by $2.16$\,s with similar spectral and temporal properties in the prompt emission, it was recently proposed that GRB~200716C might not be a long-GRB, but a short-GRB that is lensed by an intermediate mass black hole ($M_{\rm IMBH} \sim 10^5$\,M$_{\odot}$, \citealt{Wang2021, Yang2021}). According to this scenario, the optical source J1304$+$2938, could be a foreground galaxy hosting the intermediate mass black hole (IMBH) that gravitationally deforms the emission from GRB~200716C (hence, a background source).
%Studying J1304+2938 at multiple wavelengths is, therefore, crucial to shed light on the nature of this burst. 
%\textbf{Lensing...}
On the other hand, based on the analysis of its prompt emission light curve, it was recently proposed that GRB~200716C might not be a long-duration GRB, but a short-duration GRB that is lensed by an intermediate-mass black hole (IMBH; $M_{\rm IMBH} \sim 10^5$\,M$_{\odot}$; \citealt{Wang2021, Yang2021}). According to this scenario, the optical source J1304$+$2938 could be a foreground galaxy hosting the IMBH that gravitationally deforms the emission from GRB~200716C (hence, a background source).

The structure of the paper is the following.
The observations and their analysis are reported in Section~\ref{sec:obs}. We present and discuss our results in Section~\ref{sec:results} and \ref{sec:discussion}, respectively. In Section~\ref{sec:conclusions} we conclude with a brief summary. Throughout the paper we assume a standard $\Lambda$-CDM cosmology with $H_{0} = 69.32$\,km\,s$^{-1}$\,Mpc$^{-1}$, $\Omega_{\rm m}=0.286$ and $\Omega_{\Lambda}=0.714$ \citep{Hinshaw2013}. At $z=0.341$ (Section~\ref{sec:results}), 1\,arcsec corresponds to roughly 4.9\,kpc.

\section{Observations}
\label{sec:obs}

\subsection{Multi-wavelength archival data }
%$\alpha =13^{\rm h}04^{\rm m}02.616^{\rm s}$, $\delta = +29^{\circ}38^{\prime}39.02^{\prime\prime}$
We searched for J1304$+$2938 in publicly available data and surveys. Its coordinates are (J2000) $\alpha =13^{\rm h}04^{\rm m}02.371^{\rm s}$, $\delta = +29^{\circ}38^{\prime}40.66^{\prime\prime}$ \citep{Adelman2008}. This galaxy is present in catalogues produced with LOFAR at 130--170\,MHz (LOFAR J130402.62$+$293839.8, \citealt{Hardcastle2016}), the Wide-field Infrared Survey Explorer (WISE; \citealt{Wright2010}) at 3.4, 4.6, 12, and 22\,\textmu m (WISEA J130402.47$+$293839.3) and the SDSS \citep{Adelman2008} in the optical $z$, $i$, $r$, $g,$ and $u$ filters (SDSS J130402.37$+$293840.6, \citealt{D'Avanzo2020}). For these three surveys, we obtained the flux densities directly from the above references.

We also investigated the RACS at 0.89\,GHz, the Faint Images of the Radio Sky at Twenty-centimeters (FIRST; \citealt{Becker1995}), the NRAO Very Large Array Sky Survey (NVSS, \citealt{Condon1998}), and the APERture Tile  In Focus array (Apertif, \citealt{Adams2022}) imaging survey at 1.4\,GHz, and the VLASS at 3\,GHz. 
The angular resolution and the epoch of each observation are provided in Table \ref{tab:obs_arx}. At the radio wavelengths, the public observations with the highest angular resolution are those from the VLASS, with the beam size being 2.5$^{\prime\prime}$.  %, which corresponds to $\sim 12.3$ kpc at $z=0.341$%\footnote{At $z=0.348$, the luminosity distance is 1862 Mpc,  which gives a scale of 5.0\,kpc\,arcsec$^{-1}$.}.  
%For each of the radio surveys, 
We downloaded the FITS images from The Canadian Initiative for Radio Astronomy Data Analysis (CIRADA\footnote{\url{http://cutouts.cirada.ca}}) for the NRAO surveys, from the CSIRO ASKAP Science Data Archive (CASDA\footnote{\url{https://data.csiro.au/domain/casdaObservation}}) for the RACS, and from the Apertif DR1 documentation website\footnote{\url{https://www.astron.nl/telescopes/wsrt-apertif/apertif-dr1-documentation/}} for the Apertif imaging survey, and we subsequently performed Gaussian fits with the \texttt{JMFIT} task in the Astronomical Image Processing System ({\sc aips}; \citealt{Greisen2003}).
We show the radio measurements in Fig.~\ref{fig:spectrum}, while a full spectrum from 0.1 to 10$^6$\,GHz is provided in Fig.~\ref{fig:spectrum_opt} in Appendix~\ref{appendix1}.

\begin{figure}
\centering
\includegraphics[width=0.99\columnwidth]{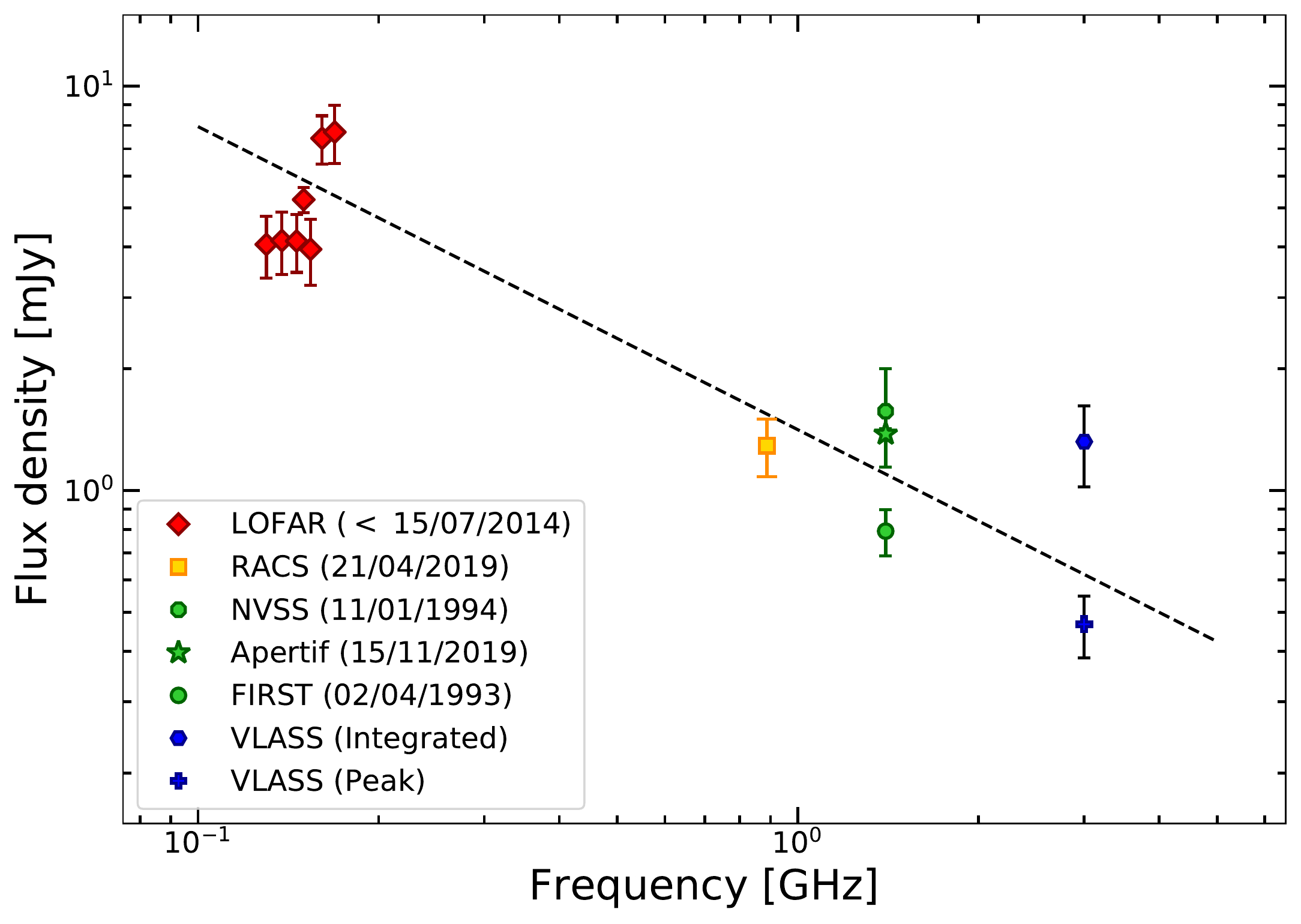}
\caption{Flux-density measurements (mJy) as a function of frequency (GHz) for J1304$+$2938 in the radio band. Data were taken at different epochs (see Table~\ref{tab:obs_arx}). The dashed black line corresponds to a power law $F\propto\nu^{\alpha}$ with spectral index $\alpha = -0.75$.}
\label{fig:spectrum}
\end{figure}

\subsection{European VLBI Network and {\it e}-MERLIN follow up}

% 5 GHz -- temporal order
We also carried out dedicated very long baseline interferometry (VLBI) observations of J1304$+$2938. On 2021 October 23, we observed at 5\,GHz with the European VLBI Network (EVN)  for a total time of 6\,h (PI: Giarratana; project: EG118A). These data were recorded at 2048\,Mbits\,s$^{-1}$ and correlated at the Joint Institute for VLBI in Europe (JIVE) into eight sub-bands (IFs) with 32\,MHz bandwidth and 64 channels each, through two polarisations (RR, LL). 
%Jb-1, Wb, Ef, Mc, Nt, On-85, Tm65, Ur, Tr, Ys, Hh, Sv, Zc, Bd, Ir, Km ANTENNAS at 5 GHz from prtan

%1 GHz -- the observations happened a week later
On 2021 October 30, we performed a sensitive 12\,h  observation with the EVN including the {enhanced} Multi-Element Remotely Linked Interferometer Network ({\it e}$-$MERLIN) at 1.6\,GHz (PI: Giarratana; project: EG118B) .
%18 antennas, Jb-1, Wb, Ef, Mc, Nt, On-85,  Ur, Tr, Hh, Sv, Zc, Bd, 
%Ir, Cm, Da, Kn, Pi, De
The data were recorded at 1024\,Mbits\,s$^{-1}$ and correlated at JIVE into eight sub-bands (IFs) with 16\,MHz bandwidth and 32 channels each, through two polarisations (RR, LL).  The averaging time for the visibilities was of 2\,s. 

The structure of the observations followed a typical phase-referencing experiment, with scans of $\sim 3$\,min on the target followed by scans of $\sim 1.5$\,min on two phase reference sources (J1310+3220 and J1300+2830). 3C345 was the fringe finder and bandpass calibrator for both the 1.6 and 5\,GHz observations.

The calibration and imaging were performed using AIPS following the standard procedure for EVN phase referenced observations\footnote{\url{https://www.evlbi.org/evn-data-reduction-guide}}, except that for the global fringe fitting, for which we used both the phase calibrators in the following way. We first derived the solutions for J1300+2830, which we applied to the target and the other calibrators. We then derived the residual solutions using a model of the other calibrator J1310+3220, and applied these final solutions to J1310+3220 and the target.
%metodo 3 dell'email di Marcello del 30 dic

The time- and bandwidth-limited field of view of these observations was of about $\sim 5$\,arcsec, but the source is well localised in the observations with an angular resolution of 2.5\,arcsec. Therefore, we searched for the radio emission of the putative host galaxy in an area of 2.5\,arcsec in diameter, which corresponds to $\sim$12\,kpc at $z = 0.341$ (see also Section~\ref{sec:results}). We adopted a natural weighting scheme to maximise the sensitivity to detect any potential extended structure. We obtained dirty images with an rms of 8\,\textmu Jy\,beam$^{-1}$ at 1.6\,GHz, and 9.6\,\textmu Jy\,beam$^{-1}$ at 5\,GHz.
At 1.6\,GHz, the largest angular scale detectable $\vartheta_{\rm LAS}$ is of about 2\,arcsec, which corresponds to roughly 10\,kpc at $z = 0.341$, while at 5\,GHz it is $\vartheta_{\rm LAS} \sim 50$\,mas, which amounts to 245\,pc.
% I find the minimum baselins of 4 Mlambda in the 5 GHz data and of about 100 klambda in the 1.6 GHz data
%We did not detect any source at $>5\sigma$ level, neither compact nor extended, at both frequencies within a field-of-view of 2.5\,arcsec. \textcolor{teal}{[Please, double check these numbers]}

\subsection{Spectroscopy from the Telescopio
Nazionale Galileo}
\label{subsec:tng}
We performed a dedicated spectroscopic follow up of J1304$+$2938 with the Device Optimized for LOw RESolution (DOLORES) installed at the Telescopio Nazionale Galileo (TNG), with the aim of confirming its photometric redshift of $z = 0.348 \pm 0.053$ as reported by the SDSS \citep{D'Avanzo2020}. We took a single 30\,min observation on the night of 2022 March 5 with the LR-B grism and a long-slit of 1.0$^{\prime\prime}$ width. The mean air mass during the observation was 1.05.

\begin{figure}
        \centering
        \includegraphics[width=0.99\columnwidth]{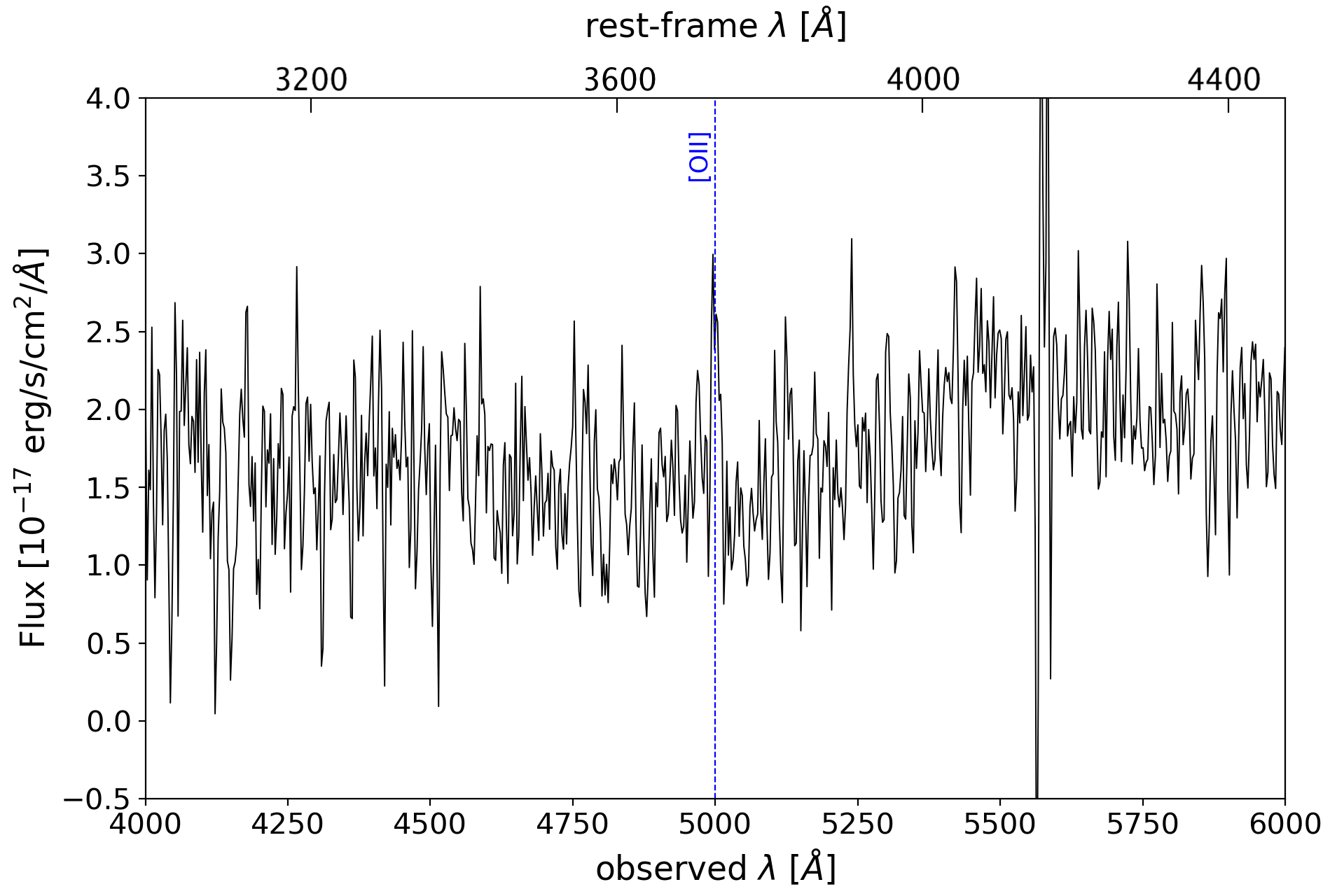}
        \caption{TNG DOLORES observed spectrum of the galaxy J1034$+$2938. The [OII]$\lambda3727$\AA~line is marked. At $\sim$5600$\AA$ a residual sky line remains after the data reduction.
         The rest frame wavelengths are shown on the upper x-axis. }
        \label{tngspec}
\end{figure}

An exposure of a He+Ne+Hg lamp was done to ensure the wavelength calibration and the flux calibration was obtained by observing the Feige~67 ($\alpha= 12^{\rm h}41^{\rm m}51.80^{\rm s}$, $\delta = +17^{\circ}31^{\prime}21.0^{\prime\prime}$) spectro-photometric standard star of the catalogs of \citet{Oke1990}. The data reduction was performed using standard Image Reduction and Analysis Facility (IRAF) procedures \citep{Tody1993}. 
The DOLORES spectrum is shown in Fig.~\ref{tngspec}.
We then smoothed the spectrum with a three-pixel boxcar to reduce the noise. Starting from the photometric redshift, we were able to identify one single emission line, corresponding to OII$\lambda$3727\AA, with a signal-to-noise ratio (S/N)\ of 10.5. The continuum is detected with a S/N of 3.2. We measured the object redshift by fitting the line with a single Gaussian profile using the IRAF task \emph{splot}. Although H$\beta$ ($\sim$6524\AA) and OIII$\lambda$5007\AA~($\sim$6720\AA) emission lines fall in the wavelength range covered by the DOLORES spectrum, we did not detect them. This could suggests that the emission of the two lines is very weak and drowned in the spectrum noise. A further observation is necessary to place any constraints on the H$\beta$ and OIII$\lambda$5007\AA~emission.
%After smoothing the spectrum with a 3 pixels boxcar to reduce the noise, we were able to measure the object redshift by fitting the only detected emission line, the OII$\lambda$3727\AA~visible in it with a single Gaussian profile, using the IRAF task \emph{splot}. aaaaa

\section{Results}
\label{sec:results}

%%% The redshift
Based on our TNG spectroscopic observations, we determine a redshift of $z = 0.341 \pm 0.004$.  This value confirms and refines the already-known photometric redshift of the galaxy \citep{D'Avanzo2020}. At $z=0.341$, the luminosity distance is 1825 Mpc, which gives a scale of 4.9\,kpc\,arcsec$^{-1}$.

%%% The radio detections
Inspection of the radio surveys, together with  measurements available in the literature, reveals unresolved radio emission at the location of the optical galaxy at a significance of between $\sim3\sigma$ and $\sim34\sigma$ in all the datasets. The resulting flux densities are reported in Table \ref{tab:obs_arx} and shown in the spectrum of Fig.~\ref{fig:spectrum}, with error bars reporting the $1\sigma$ nominal uncertainties from the fitting procedure. 

%%% Radio data as a function of frequency
The source is brightest at the lowest frequency, where the LOFAR flux densities range between 4.0 and 7.7 mJy. The spectrum is rather puzzling in this region, with a flat trend between 130 and 150\,MHz and a rise between 150 and 200\,MHz (Fig.~\ref{fig:spectrum}).  In the $\sim1$\,GHz region, the source is somewhat fainter; the most significant detection is achieved thanks to the most sensitive Apertif data ($1.38\pm0.04$\,mJy); the NVSS data indicate slightly larger values, while the highest resolution FIRST data show a slightly lower value, perhaps suggestive of the presence of some extended emission (see Figure \ref{fig:first_and_apertif}); however, the S/Ns of the NVSS and the FIRST are lower and the results could be considered overall consistent with Apertif. At 3\,GHz, the VLASS data are the only ones in which the fitting result suggests that the source is resolved, providing a
significantly larger value for the integrated flux density than the brightness surface peak.  However, J1304$+$2938 is located exactly on a side lobe of the relatively bright ($S_\mathrm{3\,GHz}=6.0\pm0.4$ mJy) and clearly extended radio source FIRST\,J130353.7+293734 (coincident with SDSS\,J130353.70+293733.1); considering this fact, the low S/N, and the ``quick look'' nature of the VLASS data, we cannot conclusively determine the nature of the detected source and consider the values for both components in our analysis. The nominal deconvolved size of the major axis of the component would be 5.6$^{\prime\prime}$, corresponding to $\sim$27\,kpc at $z=0.341$.

%%% Radio data as a function of time - and possible afterglow contribution
We further point out that the VLASS data were taken in two separate epochs, one before and one (85\,d) after the occurrence of the GRB. Nevertheless, the two measurements are in agreement with each other within the uncertainties, and so we cannot claim any contribution from the afterglow, whose flux density is constrained to be no higher than 180\,\textmu Jy.  As a matter of fact, under the reasonable assumption that the afterglow does not contribute to the second epoch emission, we also combined the two epochs in a single image, which allows us to obtain a better constrained fit, which is also reported in Table \ref{tab:obs_arx}.
% the low frequency data ($<200$\,MHz) show a  trend that cannot be easily explained by synchrotron self absorption or free-free absorption models, as there is  peculiar in 

% VLBI data and luminosities
On milliarcsecond scales, our deep VLBI observations did not detect any source at 1.6 and 5\,GHz. We can put stringent $3\,\sigma$ upper limits on the peak surface brightness of about $30$\,\textmu Jy\,beam$^{-1}$ at both frequencies, which corresponds to $\sim$9$\times 10^{28}$\,erg\,s$^{-1}$\,Hz$^{-1}$ if we use our spectroscopic redshift and adopt a reference spectral index of $\alpha=0.0$ (typical of compact components).  On the larger scales, the moderate S/N, the difference in angular resolution and observing epochs, and the still preliminary nature of the data from the latest surveys do not allow an accurate modeling of the spectrum, which will be the subject of a future study.  However, the overall trend of optically thin emission, perhaps with a hint of self-absorption at low frequency, indicates the nonthermal nature of the emission in the observed frequency range.  As a reference, in Fig.~\ref{fig:spectrum} we overlay  a $F_\nu \propto
\nu^{-0.75}$ power law on the observed data.  Using this reference value for the extended emission, and the  measurement with the highest S/N (from the Apertif imaging survey), we derive a luminosity at 1.4\,GHz of L$\simeq$(5.1$\pm$0.2)$\times10^{30}$\,erg\,s$^{-1}$\,Hz$^{-1}$.

%\subsection{GRB host galaxies}
%\label{subsec:host}
In order to discuss our results within a broader context, we carried out an extensive search for long-duration GRB host galaxies in the literature, looking in particular for previous observations in the radio band \citep{Berger2001, Berger2003, Michalowski2012, Hatsukade2012, Perley2013, Stanway2014, Stanway2015, Perley2015, Michalowski2015, Greiner2016}. We ended up with 87 galaxies: among these, only 20 are detected in the radio. Table~\ref{tab:lum}  presents the redshift, the radio (monochromatic) flux density and luminosity, and the SFR for the GRB host galaxies with a confirmed radio detection. The SFR was calculated from the observed flux density at a frequency $\nu$ according to the formula by \citet{Greiner2016}. The measured flux density of J1304$+$2938 is well above the upper limits found in the literature, making it the third-most luminous GRB host galaxy ever discovered. More generally, a radio emission above $10^{30}$\,erg\,s$^{-1}$\,Hz$^{-1}$ turned out to be rare (see Table~\ref{tab:lum}).

\section{Discussion}
\label{sec:discussion}
Having multi-frequency and multi-resolution data is an element of novelty in the study of GRB host galaxies, although it leads to a relatively complex picture. The spectrum in Figure~\ref{fig:spectrum} shows a scattered trend between 1 and 3\,GHz: the poor S/N of most detections in the surveys explains most of this scatter, although additional  factors at work could be external, such as scintillation; physical, such as a variable AGN; or instrumental, in the case of diffuse regions, due to the different angular resolutions of the surveys. As our VLBI observations do not reveal any compact emitters, we can rule out the scintillation scenario. In the following sections, we discuss the origin of the radio emission in the framework of the other two extreme cases: a radio-loud AGN versus emission from a diffuse star-forming region.
%In fact, if the emission comes from a diffuse region, one expects larger flux density values when including larger portions of it, i.e. when the beam width is larger.

\subsection{The radio-loud AGN}

The radio-to-optical luminosity ratio $R = F_\mathrm{radio} / F_\mathrm{opt}$ is a classical tool for characterising the radio loudness of an active galaxy \citep{Kellermann1989}. Considering the nearest available bands to those traditionally used to calculate $R$, we obtain for J1304$+$2938 a value of $R=53$, which is\ well into the radio-loud domain. The 1.4\,GHz radio luminosity from the Apertif imaging survey is ($5.1\times10^{23}$, in units of W\,Hz$^{-1}$) and the steep spectral index in the radio band would place J1304$+$2938 in the Fanaroff-Riley I (FRI) class \citep{Fanaroff1974}. However, the available data do not allow direct confirmation of the expected morphology for an FRI radio galaxy, with a compact core and twin jets ending in diffuse, edge-dimmed lobes or plumes. The survey data are overall compatible with the presence of some diffuse emission on scales of a few tens of kiloparsecs (kpc), as indicated by the apparently resolved nature of the VLASS image and the increase in total flux density when decreasing the resolution in the 1.4\,GHz data (from FIRST, to Apertif, and NVSS). If the total extension of the radio emission were confined within a few kpc, the source could be classified as FR0 \citep{Baldi2016} or a low-power compact source (LPC, \citealt{Giroletti2005}), which indeed represent a substantial fraction of the radio-loud population at lower redshift \citep{Baldi2018}.
%\footnote{\textbf{According to \citet{Baldi2015}, FR0 galaxies are defined as radio galaxies that are (i) associated with a red massive early type galaxy, (ii) that host a high mass black hole ($\ge$10^8\,M$_{\odot}$), (iii) that are spectroscopically classified as low excitation radio galaxy, and (iv) with a size of $\le$1-3\,kpc in the radio.}}

However, in spite of all the circumstantial support from the radio surveys, the AGN scenario lacks the ultimate signature, that is,\ the presence of an active compact core, either from high-energy data or from VLBI observations. In this sense, the stringent upper limits provided by our deep images argue against the presence of a compact core down to rather low luminosities. Therefore, our result disfavours the radio-loud AGN  scenario,  leaving only the less likely possibility of a strongly debeamed core (if the axis of the jets of the radio galaxy are seen under a large viewing angle) or of a recently switched-off nuclear activity \citep{Murgia2011}.  

At high energy, before the GRB detection by \emph{Swift}, in X-rays only the ROSAT satellite pointed towards this region of the sky between July and December 1990. No source is visible in the 0.1--2.4 keV image of the ROSAT All-Sky Survey \citep[RASS][]{Voges1999}. With PIMMS, assuming a power-law model with a photon index of 1.7, we could set only loose upper limits on the flux ($\sim$1$\times$10$^{-13}$\,erg\,cm$^{-2}$\,s$^{-1}$ in the 0.2--10\,keV band) and luminosity ($\lesssim$4$\times$10$^{43}$\,s$^{-1}$), which are not sufficient to conclude on the presence of AGN-related X-ray emission.

Future experiments able to test the debeamed AGN scenario would be a detection at high energy or a successful imaging of a radio-galaxy structure based on deeper radio data at intermediate angular resolution.  In this case, 
%As J1304$+$2938 is radio-loud, we would expect emission from the core of the galaxy; however, our VLBI observations do not show any evidence for an active nuclear component. A variable AGN can potentially explain the inconsistency between the values from FIRST and NVSS, and the non-detection in the RACS survey, because these observations are not simultaneous (Table \ref{tab:obs}). However, all together the archival and new VLBI upper limits produce a spectrum that is difficult to reconcile with the typical emission models for AGNs. Moreover, the 1.4 -- 3\,GHz part of the spectrum yields a spectral index of $\langle\alpha\rangle = -1.3 \pm 0.2$ (considering both the NVSS and FIRST measurements). This is a quite steep spectral index for typical radio-loud AGNs \citep[e.g.,][]{Hovatta2014}. Nevertheless, in the unlikely case that J1304+2938 hosts a variable AGN 
and if GRB~200716C belongs to J1304$+$2938, this would be the third GRB found within a galaxy with an AGN, after GRB~170817A, which occurred in NGC 4993 \citep{Coulter2017, Palmese2017, Fong2017, Contini2018, Wu2018}, and GRB~150101B, which belonged to WISEA J123204.97-105600.6 \citep{Xie2016}. NGC 4993 is a low-luminosity, radio-loud galaxy \citep{Wu2018}, while WISEA J123204.97-105600.6 (2MASX J12320498-1056010) is an X-ray bright, radio-loud galaxy \citep{Xie2016}.

\subsection{The extreme star formation}

An immediate implication of the non-detection with the EVN is that the radio emission detected by lower angular resolution surveys is consistent with being extended on scales that are larger than the largest detectable angular scale $\vartheta_{\rm LAS}$, which is of $2$\,arcsec at 1.6\,GHz (hence smaller than the angular scales sampled by the VLASS).
Moreover, the lack of a compact component disfavours variability as the most viable explanation for the discrepancy between low angular resolution measurements.  On the other hand, the trend of increasing total flux density when considering lower resolutions in the survey data at 1.4\,GHz  corroborates the hypothesis of the presence of diffuse emission on galactic scales. Moreover, considering the FIRST and the Apertif imaging surveys, the beam area is roughly 23 and 309\,arcsec$^2$ (Figure \ref{fig:first_and_apertif}), respectively, while the flux density is $(790\pm100)$ and $(1380\pm40)$\,$\upmu$Jy, respectively. Thus, in the FIRST survey we would have a contribution from the galaxy of $(590\pm108)$ \,$\upmu$Jy spread over 39 beams, and hence an average of $(15\pm3)$\,$\upmu$Jy\,beam$^{-1}$, which is under its rms noise level. This is a rather simplified approach, assuming uniform brightness distribution over the entire Apertif beam area, but it is generally in agreement with the presence of more intense star formation in the central regions (within the $\sim25$ kpc beam of the FIRST) and lower, yet significant additional regions falling in the 140\,kpc $\times$ 54\,kpc beam of Apertif.

%\textbf{Moreover, as stated above, the discrepancy between the integrated flux density and the brightness surface peak in the VLASS data suggests the presence of an extended emission. The resolution of the VLASS is 2.5\,arcsec, which corresponds to roughly 12 kpc at $z = 0.341$. If the emission is due to, e.g., a star-forming spiral galaxy, we expect it to be more extended (CITE???): that being the case, the VLASS would take into account only a portion of it. On the other hand, the Apertif imaging survey has a resolution of (28.6$\times$11.1)\,arcsec$^2$, corresponding to (140$\times$54)\,kpc$^2$, and therefore we would expect it to include the whole contribution of the galaxy.} 

We further note the presence of a second emitting component in the FIRST and Apertif imaging surveys (Figure \ref{fig:first_and_apertif}): this contaminating source is found at a distance of $\sim$40\,arcsec, which is 200\,kpc at $z = 0.341$, and is therefore likely unrelated to J1304$+$2938. However, in the NVSS, J1304$+$2938 and the contaminating source are not well separated, possibly explaining the observed discrepancy in the total flux density between the Apertif imaging survey and the NVSS.
%Furthermore, the different flux densities from the radio surveys could be due to the significantly different angular resolutions (Table \ref{tab:obs}), which sample different scales and, hence, may resolve out part of the diffuse emission of J1304+2938. For instance, the NVSS beam includes the emission within $\sim220$\,kpc, while the FIRST beam measures the emission within $\sim25$\,kpc.

Possible mechanisms for a diffuse radio emission unrelated to nuclear activity are the free-free emission from the ionised gas surrounding a population of bright OB stars, which would lead to a thermal spectrum, and/or the SN contribution from young stars, which is characterised by a steep non-thermal spectrum. As our data are clearly suggestive of a steep spectral index, we can assume the latter to be the predominant emission mechanism in the portion of the spectrum we are interested in. Considering the high luminosity we find, this leads to a high SFR\footnote{we consider a galaxy as highly star forming if SFR $\ge 15$\,M$_{\odot}$\,yr$^{-1}$ \citep{Greiner2016}}. As the SFR can be inferred from the radio luminosity with different formulas, from the flux density at 1.4\,GHz with the FIRST, the Apertif imaging survey, and the NVSS, we estimate that SFR$= (186 \pm 42)$\,M$_{\odot}$\,yr$^{-1}$, $(324 \pm 61)$\,M$_{\odot}$\,yr$^{-1}$, and $(376 \pm 117)$\,M$_{\odot}$\,yr$^{-1}$, using the conversion from \citet{Greiner2016}, respectively.
%Considering the luminosity we found with the FIRST and the NVSS surveys, from \cite{Condon1992} we reckon that the formation rate of stars with masses larger than 5\,M$_{\odot}$ is SFR(M$\ge$5\,M$_{\odot}$)=(65$\pm$10)\,M$_{\odot}$\,yr$^{-1}$ and SFR(M$\ge$5\,M$_{\odot}$)=(182$\pm$55)\,M$_{\odot}$\,yr$^{-1}$, respectively. These values are in agreement with what expected for the environments of long-GRBs \citep{Greiner2016, Klose2019}.
Even taking the more conservative SFR derived with the FIRST, J1304$+$2938 would be among the ten most-star-forming GRB host galaxies discovered so far.
%, after the host galaxies of GRB~021211 \citep{Michalowski2012, Greiner2016}, GRB~060814 \citep{Perley2015, Greiner2016}, GRB~980703 \citep{Berger2001, Greiner2016}, GRB~000418 \citep{Berger2003, Greiner2016}, GRB~100621A, GRB~050223 \citep{Stanway2014, Greiner2016}, GRB~080207 \citep{Perley2013, Greiner2016}, GRB~090404 \citep{Perley2013}.

%As the SFR derived from the radio is not affected by dust extinction, ..., which is important to characterise the environment of the GRB progenitors (CITE) and to ... dark GRBs\footnote{The so-called dark GRBs are a sub-population of GRB with little or absent optical emission \citep{Groot1998}. A possible explanation is that the GRB outflow encounters a large amount of dust along the line of sight within the host.} \citep{Hatsukade2012, Greiner2016}.
%dust attenuation important for dust amount = environment of the progenitor AND dark grb (intrinsic or due to dust?)
As the SFR derived from the radio is not affected by dust extinction, by comparing it with the value provided by optical estimators, it is possible to determine the amount of dust within the host galaxy, which is important for characterising the environment that leads to a burst \citep{Berger2001, Berger2003}. To obtain meaningful constraints on the SFR, \citet{Michalowski2012} used a complete sample of 30 hosts with $z < 1$, including those from the The Optically Unbiased Gamma-Ray Burst Host (TOUGH) sample \citep{Hjorth2012} and sources compiled from the literature. The authors found that at least $\sim$63\% of GRB hosts have SFR $<$ 100\,M$_{\odot}$\,yr$^{-1}$ and at most $\sim$8\% can have SFR $>$ 500\,M$_{\odot}$\,yr$^{-1}$. Surprisingly, $\gtrsim$ 88\% of the $z \lesssim 1$ GRB hosts have UV dust attenuation A$_{UV}$ $<$ 6.7\,mag and A$_{V}$ $<$ 3\,mag, suggesting that the majority of GRB host galaxies are not heavily obscured by dust. The latter result is further strengthened by subsequent studies on samples of GRB hosts \citep[see e.g. ][]{Hatsukade2012, Perley2013, Stanway2014, Greiner2016}. %\citet{Perley2015} studied a sample of 32 hosts with $0 < z < 2.5$ and found that between 9\% and 23\% of GRBs with $0.5 < z < 2.5$ occur in galaxies with SFR $>$ 50\,M$_{\odot}$\,yr$^{-1}$ at $z \sim 1$, or $>$250\,M$_{\odot}$\,yr$^{-1}$ at $z \sim 2$.
%Complementary studies on dark GRB\footnote{The so-called dark GRBs are a sub-population of GRB with little or absent optical emission \citep{Groot1998}. A possible explanation is that the GRB outflow encounters a large amount of dust along the line of sight within the host.} host galaxies at different redshifts ($z \le 1.006$ \citealt{Hatsukade2012}; $z \le 3.038$ \citealt{Perley2013}; $z \le 1.9$ \citealt{Greiner2016}) did not reveal extreme SFRs from radio observations, further corroborating the hypothesis that the dust obscuration is negligible for most of these sources, and that dark GRBs do not always occur in galaxies enshrouded by dust, with direct implications on the possible intrinsic faintness of the optical afterglow \citep{Hatsukade2012}.
To determine the level of dust obscuration, a reliable estimate of the SFR from optical estimators is needed, such as the H$\alpha$, H$\beta,$ or NII emission lines, which could also provide further confirmation of the photometric redshift. Above all, such optical estimators would allow a detailed study of the chemical composition of the galaxy. Among these, the H$\beta$ %and OIII$\lambda$5007\AA~
emission line falls in the wavelength range covered; nevertheless, our spectral observation does not allow us to calculate the SFR from the latter emission lines, possibly due to the fact that they are too weak. A preliminary estimate of the flux expected from the H$\beta$ emission line can be provided by taking the relation between OII, H$\alpha,$ and H$\beta$ from \citet{Argence2009}, and assuming the ratio OII/H$\alpha = 1.26$ for star-forming galaxies in the  local Universe provided by \citet{Mouhcine2005}; for a S/N of 10.5 for the OII detection, we get a flux three times smaller for the H$\beta$ emission line, which would be too weak to be detected above the continuum emission. %Moreover, although it has been suggested that the OII$\lambda$3727\AA~forbidden line is a good substitute in high-redshift galaxies after a proper calibration using the H$\alpha$ \citep[see, e.g., ][]{Gallagher1989, Kennicutt1992, Kennicutt1998}, the conversion from OII$\lambda$3727\AA~ luminosity to SFR can be affected by the effects of metallicity \citep{Talia2015}. 
Further, deeper spectroscopic follow up is therefore needed.%: in fact, if the SFR derived from the optical is order of magnitudes lower than that derived from the radio, we could reveal an outstanding source, whose properties would not follow the general trend pictured so far.
%In order to more securely estimate the contribution of the diffuse star formation within J1304$+$2938 a deeper spectroscopic follow-up in the optical is necessary. The detection of typical emission lines of star-forming galaxies, such as H$\alpha$ or NII, could provide a robust independent estimate of the SFR and also allow one to study in detail the chemical composition of the galaxy.

\begin{table*}[t]
\centering
\begin{threeparttable}
\centering
\caption[]{VLBI observations of J1304$+$2938.}
\begin{tabular}{ccccc}
\toprule
Array          &Central frequency    &T-T$_0$    &Angular resolution    & Flux density\\
        &(GHz)         & (days)     & (arcsec)    & (\textmu Jy)\\
\midrule
EVN+{\it e}$-$MERLIN &1.6 & 379 & 0.010 & $<24$\\
EVN & 5.0  & 372 & 0.005 & $<29$\\
\bottomrule
\end{tabular}
\label{tab:obs}
\begin{tablenotes}
\item \textsl{Column 1:} Array; \textsl{Column 2:} observing frequency (GHz); \textsl{Column 3:}  
T-T$_0$ (days), which is the total time from the burst; \textsl{Column 4:} angular resolution (arcsec); \textsl{Column 5:} Upper limits (3\,$\sigma$) for the flux density (\textmu Jy).
\end{tablenotes}
\end{threeparttable}
\end{table*}

\begin{figure}
\centering
\includegraphics[width=0.48\textwidth]{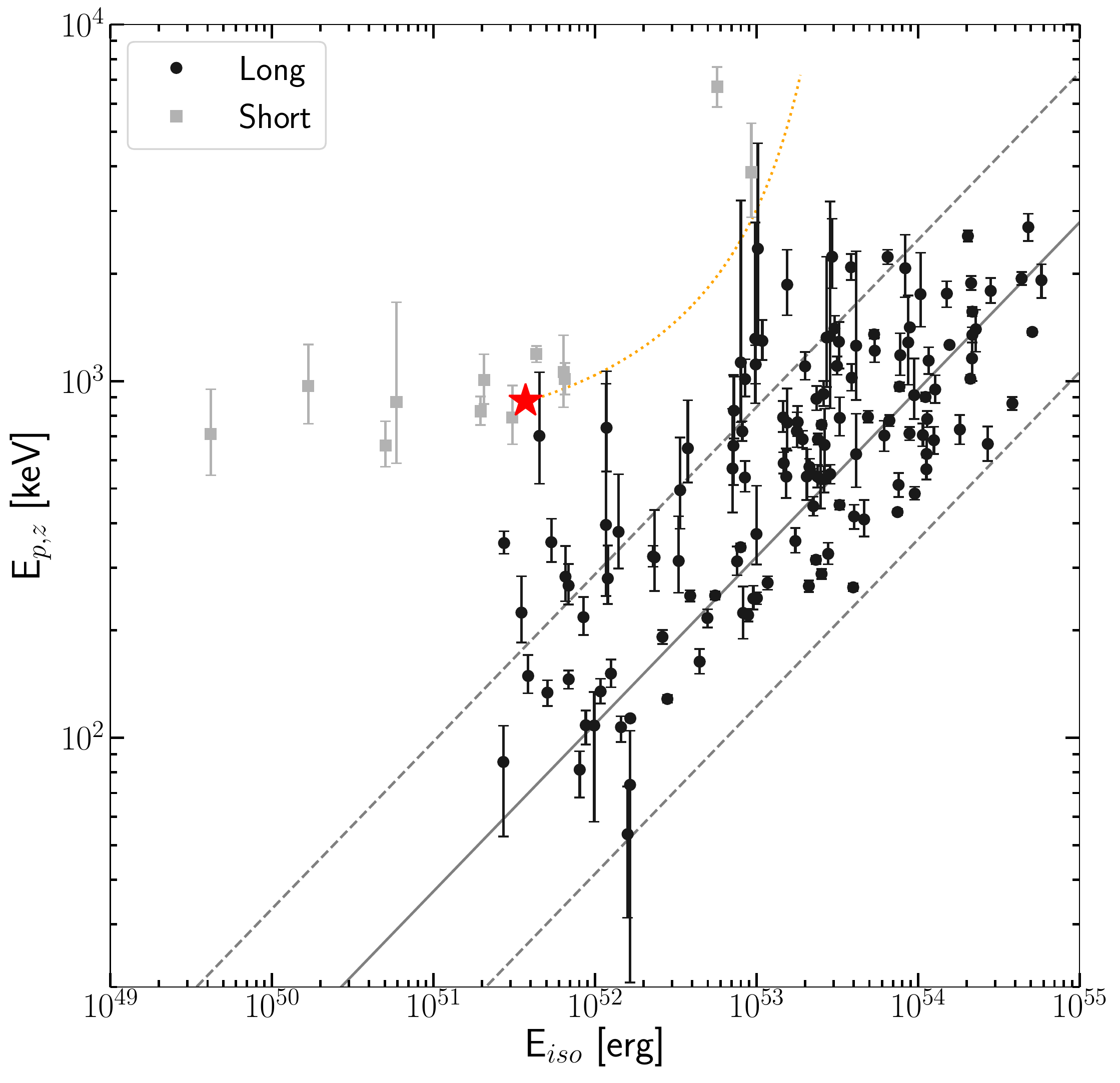}
\caption{Location of GRB~200716 (red star) in the rest-frame (E$_{iso}$, E$_{p,z}$) plane for the short-duration GRBs (grey squares) and the long-duration GRBs (black circles) of \citet{Tsvetkova2017}. The grey solid line indicates the Amati relation estimated using the long-duration GRBs of \citet{Tsvetkova2017}, while the grey dashed lines indicate its 3$\sigma$ uncertainty. The orange dotted line shows the position of the burst for 0.341 $\le$ z $\le$ 10.}
\label{fig:amati_relation}
\end{figure}

\subsection{J1304+2938 and GRB 200716C}
\label{subsec:GRBandHost}
The overall radio properties of J1304$+$2938 seem to favour a highly star-forming galaxy, which is the natural environment expected for explosive transient events generated during the collapse of young massive stars, such as long-duration GRBs. Therefore, the radio properties of J1304$+$2938 are in agreement with the long-duration nature of GRB~200716C.
Nevertheless, there are still some caveats that are relevant to the interpretation of this burst. First of all, the spectrum of the galaxy in the radio band shows some peculiarities that could be due to the low S/N, the different angular resolutions, and/or the epochs of the surveys. To solve the conundrum, deep observations with arcsecond resolution and a broad bandwidth are required, such those provided by the Karl G. Jansky Very Large Array, for example.
Second, taking the isotropic equivalent energy E$_{iso}$ and the time-integrated peak energy E$_{p}$ for 150 long- and short-duration Konus/Wind GRBs \citep{Tsvetkova2017}, with E$_{iso} = 3.7\times10^{51}$\,erg and E$_{p,z} = $880\,keV, GRB~200716C is a clear outlier of the Amati relation, where E$_{p,z} =$ E$_{p}(1+z)$ (see Fig.~\ref{fig:amati_relation}) and E$_{iso}$ was rescaled to $z=0.341$. %Even if J1304$+$2938 is a foreground galaxy with GRB~200716C being a background source, i.e. at higher $z$, this GRB will still not lie in the long-GRB area 
This holds true even in the case where J1304$+$2938 is a foreground galaxy and GRB~200716C is at a higher redshift (Fig.~\ref{fig:amati_relation}, orange dotted line). To be consistent with the 3\,$\sigma$ uncertainty of the Amati relation, the uncertainty on the peak energy should be at least $\sim$230\,keV (1\,$\sigma$). Finally, we note that GRB~200716C is located close to another well-known and still puzzling outlier of the E$_{iso}$-E$_{p,z}$ relation, namely GRB~061021 \citep{Nava2012}. Nevertheless, the GRB~061021 host galaxy was not detected up to 6\,$\upmu$Jy\,beam$^{-1}$ at 6\,GHz \citep{Eftekhari2021}, suggesting different properties with respect to J1304$+$2938.

The position of GRB~200716C on the E$_{iso}$--E$_{p,z}$ plane, together with the fact that
its prompt emission light curve shows two prominent peaks, followed by an extended emission up to T$_{90} \sim$90\,s \citep{Veres2020, Barthelmy2020, Torii2020, Xue2020}, led some authors to question the long-duration nature of this burst.
An alternative explanation could be that GRB~200716C is a short-duration GRB gravitationally lensed by an IMBH, which is probably hosted by J1304$+$2938 \citep{Wang2021, Yang2021}.
%\textbf{In fact,} from the analysis of the $\gamma$-ray light curve \citet{Wang2021} and \citet{Yang2021} suggested that GRB~200716C might be a short-GRB gravitationally lensed by an IMBH probably hosted by J1304$+$2938. 
We highlight the fact that, because of their high luminosities (up to 10$^{53-54}$\,erg\,s$^{-1}$; \citealt{Piran2004, Kumar2015}), GRBs can be detected up to the highest refshifts (the farthest GRB currently known is GRB~090429B at a photometric redshift of $z = 9.4$; \citealt{Cucchiara2011}) and therefore they can be used as probes of the early Universe \citep{Fryer2021}. As they could be cosmologically distant events, some GRBs might be gravitationally lensed (e.g. \citealt{Paynter2021} and references therein). Because of the strong lensing effect, photons coming from a distant source travel different geometric paths as they approach the foreground lensing object and form multiple magnified images of the same background source \citep{Congdon2018}. As a consequence, we observe variations in the lensed images with a time delay that depends on the gravitational potential of the lens. In the case of GRBs, if gravitationally lensed, we expect to measure a bright $\gamma$-ray pulse followed by a dimmer duplicate. To date only a few GRBs have been suggested as candidate lensed events, namely GRB~950830 \citep{Paynter2021}, GRB~210812A \citep{Veres2021}, GRB~081126A, and GRB~090717A \citep{Lin2021}, based on the analysis of their light curves.

If J1304$+$2938 hosts the gravitational lens of GRB~200716C, VLBI observations could potentially detect a compact emission from a radio-loud IMBH acting as a (milli-)lens \citep[e.g.][]{Paragi2006}. Possible radio emission from an IMBH would greatly help our understanding of the localisation of these objects in galaxies, which is highly unconstrained from an observational perspective \citep[e.g.][]{Weller2022}. Ultra-luminous X-ray sources (ULXs) have been suggested as possible IMBHs \citep{Kaaret2001, Miller2003} and they are variable objects on different timescales (from months to years; see e.g. \citealt{Lasota2011, Earnshaw2016, Atapin2019}). However, not even our sensitive VLBI follow up can shed light on this hypothesis as the radio emission from accreting IMBH can only be detected in local galaxies \citep{Cseh2015,Mezcua2018}.

To date, only a few (macro-)lensing galaxies showing radio/mm emission \citep{McKean2007, Haas2014, Paraficz2018} have been found, making `radio-emitting' lenses extremely rare objects\footnote{These radio-loud lenses are at higher redshifts than J1304$+$2938 ($z\sim0.65-0.8$).}.
In general, VLBI is the only method that allows us to pinpoint the multiple images produced by a gravitational lens with mass $<10^{5-6}$\,M$_{\odot}$, which are expected to be separated by a few mas \citep{Spingola2019, Casadio2021}. Nevertheless, in order to detect the putative radio-lensed images of GRB~200716C, the VLBI observations would have had to be carried out within a few hours or days of the detection of the burst at $\gamma$-rays.

\section{Conclusions}
\label{sec:conclusions}
In this paper, we present the analysis of dedicated VLBI observations together with IR and optical public data of the putative host galaxy J1304$+$2938 of GRB~200716C at $z=0.341$. We set stringent upper limits (sensitivity of $<10$\,\textmu Jy\,beam$^{-1}$) on the presence of compact radio emission, namely $<$50\,mas at 5\,GHz, within a field of view of 2.5 arcsec at 1.6 and 5\,GHz. Moreover, by performing a dedicated spectroscopic follow up with the TNG, we corroborate the previous redshift estimate of the galaxy \citep{D'Avanzo2020}. 

The non-detection with EVN and EVN+{\it e}-MERLIN suggests that the radio emission detected at low angular resolution by the RACS, FIRST, the Apertif imaging survey, and the NVSS and VLASS surveys might be diffuse and therefore completely resolved out by our VLBI observations. Moreover, the observed scatter in the publicly available flux density measurements at low frequencies cannot be explained by a variable, compact source, further corroborating the hypothesis of diffuse emission from highly star-forming regions. We derive a 1.4\,GHz luminosity of greater than $10^{30}$\,erg\,s$^{-1}$\,Hz$^{-1}$, which implies a SFR $\sim300$\,M$_{\odot}$\,yr$^{-1}$. This high SFR is consistent with the most extreme environments for long-duration GRBs. %, favouring a scenario of an unlensed long-GRB hosted by J1304$+$2938. 
That being the case, J1304$+$2938 would be among the most  radio-bright long-GRB host galaxies discovered so far. Nevertheless, the temporal and spectral properties of the prompt emission of GRB~200716C, together with the offset with respect to the Amati relation for long-duration GRBs, mean that the nature of this burst remains puzzling.
%
%would favour a scenario of an unlensed long-GRB hosted by J1304+2938, but for being a long-GRB is strangely offset with respect to the Amati relation.

\begin{acknowledgements}
We thank the referee for their useful suggestions and comments.
%instruments
The European VLBI Network is a joint facility of independent European, African, Asian, and North American radio astronomy institutes. Scientific results from data presented in this publication are derived from the following EVN project codes: EG118A, EG118B. e-MERLIN is a National Facility operated by the University of Manchester at Jodrell Bank Observatory on behalf of STFC.

This work is based on observations made with the Italian Telescopio Nazionale Galileo (TNG) operated on the island of La Palma by the Fundación Galileo Galilei of the INAF (Istituto Nazionale di Astrofisica) at the Spanish Observatorio del Roque de los Muchachos of the Instituto de Astrofisica de Canarias. The observations were executed by M. Pedani on a night with a short slot of DDT time available.

This research has made use of the CIRADA cutout service at URL cutouts.cirada.ca, operated by the Canadian Initiative for Radio Astronomy Data Analysis (CIRADA). CIRADA is funded by a grant from the Canada Foundation for Innovation 2017 Innovation Fund (Project 35999), as well as by the Provinces of Ontario, British Columbia, Alberta, Manitoba and Quebec, in collaboration with the National Research Council of Canada, the US National Radio Astronomy Observatory and Australia’s Commonwealth Scientific and Industrial Research Organisation.This paper includes archived data obtained through the CSIRO ASKAP Science Data Archive, CASDA (http://data.csiro.au). This publication makes use of data products from the Wide-field Infrared Survey Explorer, which is a joint project of the University of California, Los Angeles, and the Jet Propulsion Laboratory/California Institute of Technology, funded by the National Aeronautics and Space Administration.

Funding for the Sloan Digital Sky Survey IV has been provided by the Alfred P. Sloan Foundation, the U.S. Department of Energy Office of Science, and the Participating Institutions. SDSS-IV acknowledges support and resources from the Center for High Performance Computing  at the University of Utah. The SDSS website is www.sdss.org. SDSS-IV is managed by the Astrophysical Research Consortium for the Participating Institutions of the SDSS Collaboration including the Brazilian Participation Group, the Carnegie Institution for Science, Carnegie Mellon University, Center for Astrophysics | Harvard \& Smithsonian, the Chilean Participation Group, the French Participation Group, Instituto de Astrof\'isica de Canarias, The Johns Hopkins University, Kavli Institute for the Physics and Mathematics of the Universe (IPMU) / University of Tokyo, the Korean Participation Group, Lawrence Berkeley National Laboratory, Leibniz Institut f\"ur Astrophysik Potsdam (AIP),  Max-Planck-Institut f\"ur Astronomie (MPIA Heidelberg), Max-Planck-Institut f\"ur Astrophysik (MPA Garching), Max-Planck-Institut f\"ur Extraterrestrische Physik (MPE), National Astronomical Observatories of China, New Mexico State University, New York University, University of Notre Dame, Observat\'ario Nacional / MCTI, The Ohio State University, Pennsylvania State University, Shanghai Astronomical Observatory, United Kingdom Participation Group, Universidad Nacional Aut\'onoma de M\'exico, University of Arizona, University of Colorado Boulder, University of Oxford, University of Portsmouth, University of Utah, University of Virginia, University of Washington, University of Wisconsin, Vanderbilt University, and Yale University.

This work makes use of data from the Apertif system installed at the Westerbork Synthesis Radio Telescope owned by ASTRON. ASTRON, the Netherlands Institute for Radio Astronomy, is an institute of the Dutch Research Council (“De Nederlandse Organisatie voor Wetenschappelijk Onderzoek, NWO).

%people
The authors thank the directors and staff of all the EVN telescopes for making the observations possible.

%funding
CS acknowledges financial support from the Italian Ministry
of University and Research $-$ Project Proposal CIR01$\_$00010.

\end{acknowledgements}

% CS: mi è sembrato che l'altro modo di inserire la bibliografia non facesse distinzione tra ciò che effettivamente è citato e ciò che invece non lo è

\bibliographystyle{aa} % style aa.bst
\bibliography{main} % your references references.bib

\begin{appendix}
\onecolumn
\section{Photometric data}
\label{appendix1}

\begin{figure}
\centering
\includegraphics[width=0.5\columnwidth]{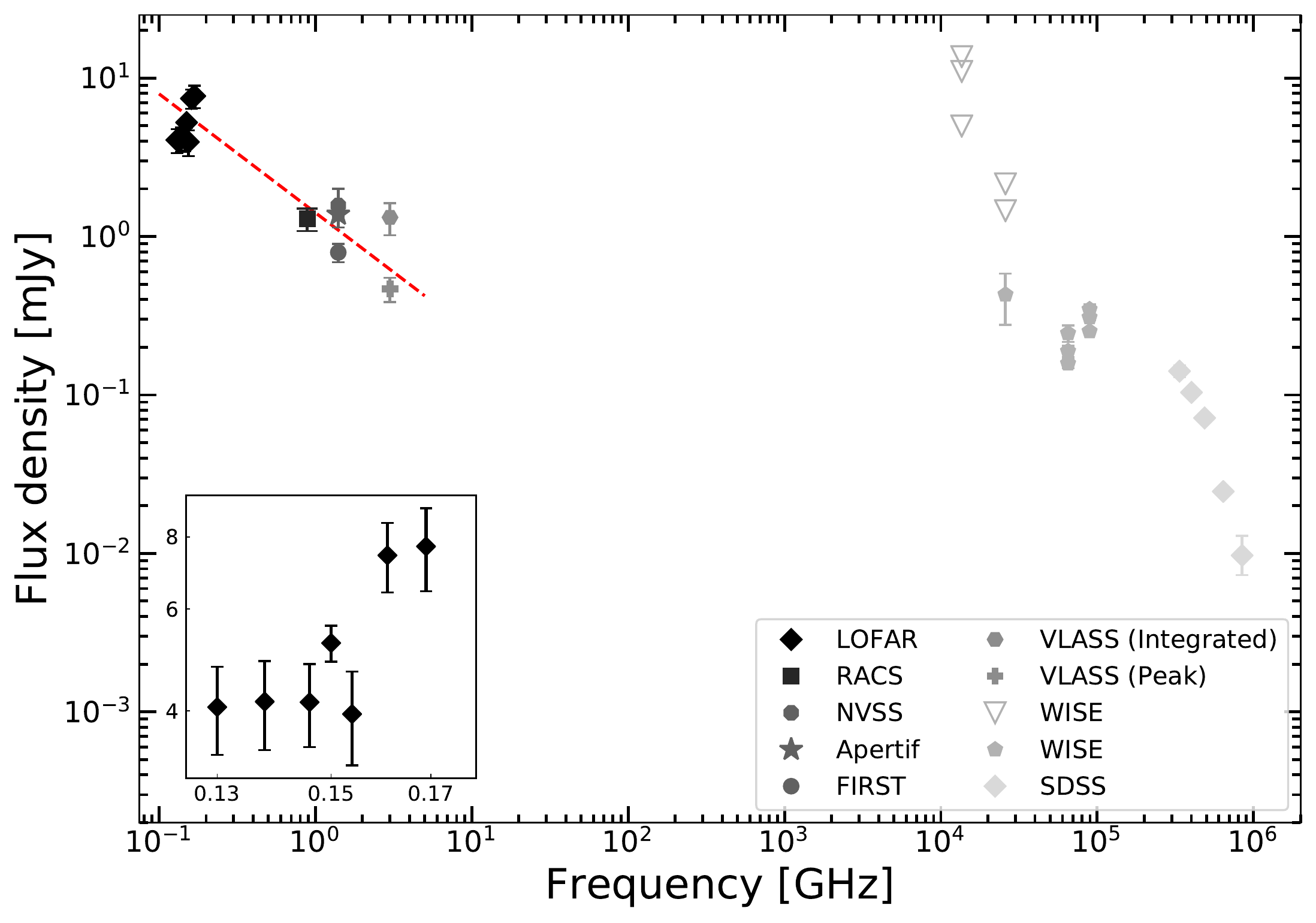}
\caption[]{Flux-density measurements (mJy) as a function of frequency (GHz) for J1304$+$2938 from 0.1 to 10$^6$\,GHz. The inset shows the LOFAR data, while the arrows indicate the 3$\sigma$ upper limits. Data are taken at different epochs (see Table~\ref{tab:obs_arx}). The dashed red line corresponds to a power law $F\propto\nu^{\alpha}$ with spectral index $\alpha = -0.75$.}
\label{fig:spectrum_opt}
\end{figure}

\begin{figure}
\centering
\includegraphics[width=0.6\columnwidth]{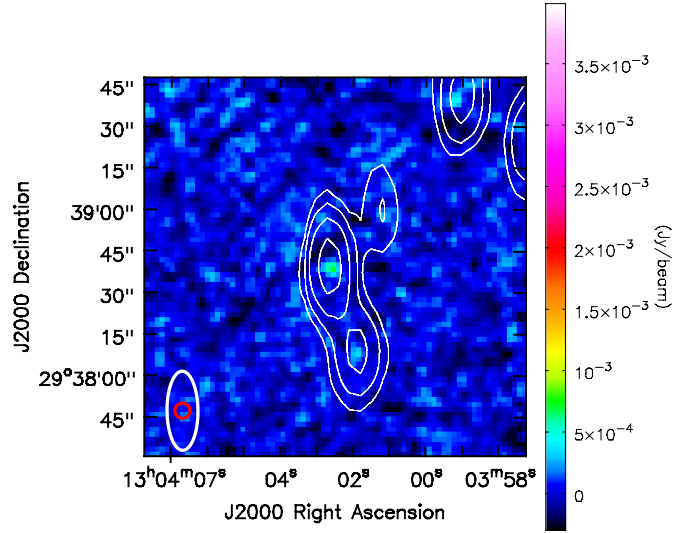}
\caption[]{Radio detection of J1304$+$2938 in the FIRST survey at 1.4\,GHz, shown by the coloured map and the associated colour bar. The surface brightness contours at levels of 3, 6, 12, 24, and 48$\sigma$  from the Apertif imaging survey are superimposed in white, where the rms noise level of the Apertif imaging survey is $\sigma = 40\,\upmu$Jy\,beam$^{-1}$. On the lower left, the restoring beams are shown as a red and a white ellipse for the FIRST and the Apertif imaging survey, respectively. A second, resolved source at roughly 40\,arcsec is found to the south.}
\label{fig:first_and_apertif}
\end{figure}

Table \ref{tab:obs_arx} presents the various measurements for J1304$+$2938 available from the literature and/or our analysis of survey data. Figure \ref{fig:spectrum_opt}  presents the flux density measurements (mJy) as a function of frequency (GHz), from 0.1 to 10$^6$\,GHz. Figure \ref{fig:first_and_apertif} shows the radio detection of J1304$+$2938 in the FIRST (colours) and Apertif imaging survey (surface brightness contours, in white).

\begin{table}[h]
\centering
\begin{threeparttable}
%\resizebox{cm}{!}{
\centering
\caption[]{Publicly available data for J1304$+$2938 from different surveys.}
\begin{tabular}{cccccc}
\toprule
Survey / Instrument          &Central frequency    &Date    &Angular resolution    & Flux density   &Ref.\\
        &(GHz)         &     & (arcsec)    & (mJy)   &\\
\midrule
LOFAR &0.130  &$<$15/07/2014   &6$\times$10 &4.1$\pm$0.7   &\citet{Hardcastle2016}\\
LOFAR &0.138  &$<$15/07/2014   &6$\times$10 &4.2$\pm$0.7   &\citet{Hardcastle2016}\\
LOFAR &0.146  &$<$15/07/2014   &6$\times$10 &4.1$\pm$0.7   &\citet{Hardcastle2016}\\
LOFAR &0.150  &$<$15/07/2014  &6$\times$10 &5.2$\pm$0.4   &\citet{Hardcastle2016}\\
LOFAR &0.154  &$<$15/07/2014   &6$\times$10 &4.0$\pm$0.7   &\citet{Hardcastle2016}\\
LOFAR &0.161  &$<$15/07/2014   &6$\times$10 &7.4$\pm$1.0   &\citet{Hardcastle2016}\\
LOFAR &0.169  &$<$15/07/2014   &6$\times$10 &7.7$\pm$1.3   &\citet{Hardcastle2016}\\
RACS  &0.89   &21/04/2019 &26$\times$11       &1.3$\pm$0.2   &This work\\
FIRST &1.4  &02/04/1993 &5.4  &0.79$\pm$0.10   &This work\\
Apertif imaging survey &1.4  &15/11/2019 &28.6$\times$11.1  &1.38$\pm$0.04   &This work\\
NVSS  &1.4  &11/01/1994 &45           &1.6$\pm$0.4   &This work\\
VLASS1\tnote{a} &3  &25/11/2017 &3.0$\times$2.3        &0.38$\pm$0.10   &This work\\
VLASS1\tnote{b} &3  &25/11/2017 &3.0$\times$2.3        &1.4$\pm$0.5   &This work\\
VLASS2\tnote{a} &3  &9/10/2020 &2.7$\times$2.4         &0.56$\pm$0.12   &This work\\
VLASS2\tnote{b} &3  &9/10/2020 &2.7$\times$2.4         &1.3$\pm$0.3   &This work\\
VLASS-combined\tnote{a} &3  & &3.0$\times$2.3         &0.47$\pm$0.08   &This work\\
VLASS-combined\tnote{b} &3  & &3.0$\times$2.3         &1.3$\pm$0.3   &This work\\
WISE &1.36$\times10^4$  &$<$06/08/2010 &12  &$<1.6$   &\citet{Wright2010}\\
WISE &1.36$\times10^4$  &$<$06/08/2010 &12  &$<3.7$   &\citet{Wright2010}\\
WISE &1.36$\times10^4$  &$<$06/08/2010 &12  & $<4.6$   &\citet{Wright2010}\\
WISE &2.59$\times10^4$  &$<$06/08/2010 &6.5 & $<0.5$   &\citet{Wright2010}\\
WISE &2.59$\times10^4$  &$<$06/08/2010 &6.5 & $<0.7$   &\citet{Wright2010}\\
WISE &2.59$\times10^4$  &$<$06/08/2010 &6.5 &0.4$\pm$0.2   &\citet{Wright2010}\\
WISE &6.51$\times10^4$  &$<$06/08/2010 &6.4 &0.158$\pm$0.011   &\citet{Wright2010}\\
WISE &6.51$\times10^4$  &$<$06/08/2010 &6.4 &0.19$\pm$0.02   &\citet{Wright2010}\\
WISE &6.51$\times10^4$  &$<$06/08/2010 &6.4 &0.24$\pm$0.03   &\citet{Wright2010}\\
WISE &8.94$\times10^4$  &$<$06/08/2010 &6.1 &0.252$\pm$0.008   &\citet{Wright2010}\\
WISE &8.94$\times10^4$  &$<$06/08/2010 &6.1 &0.309$\pm$0.012   &\citet{Wright2010}\\
WISE &8.94$\times10^4$  &$<$06/08/2010 &6.1 &0.35$\pm$0.03   &\citet{Wright2010}\\
SDSS ($z$) &3.36$\times10^5$  &23/05/2004 &         &0.141$\pm$0.012   &\citet{Ahumada2020}\\
SDSS ($i$) &4.01$\times10^5$  &23/05/2004 &         &0.104$\pm$0.003   &\citet{Ahumada2020}\\
SDSS ($r$) &4.86$\times10^5$  &23/05/2004 &         &0.072$\pm$0.002   &\citet{Ahumada2020}\\
SDSS ($g$) &6.40$\times10^5$  &23/05/2004 &         &0.024$\pm$0.001   &\citet{Ahumada2020}\\
SDSS ($u$) &8.45$\times10^5$  &23/05/2004 &         &0.010$\pm$0.003   &\citet{Ahumada2020}\\
\bottomrule
\end{tabular}
\label{tab:obs_arx}
\begin{tablenotes}
\item \textsl{Column 1:} survey or instrument. \textsl{Column 2:} observing frequency (GHz). \textsl{Column 3:} Date of the observation. \textsl{Column 4:} angular resolution (arcsec). \textsl{Column 5:} Flux density (mJy). The upper limits for the flux density are given with a 1\,$\sigma$ confidence. \textsl{Column 6:} References.
\item [a] From JMFIT peak intensity.
\item [b] From JMFIT integral intensity.
%\item [c] Brightness surface peak for the average of the two epochs.
%\item [d] Integrated flux density for the average of the two epochs.
\end{tablenotes}
\end{threeparttable}
\end{table}

\clearpage

\section{Luminosities of the GRB host galaxies}
\label{appendix2}
Table \ref{tab:lum}  presents the redshift, the radio (monochromatic) luminosity, the frequency, and star-formation rate (SFR) for the GRB host galaxies detected in radio. The SFR was calculated from the observed flux density at a frequency $\nu$ according to the following formula \citep{Greiner2016}:
\begin{equation}
    \Biggl(\frac{\mathrm{SFR}}{\mathrm{M}_{\odot}/\mathrm{yr}}\Biggr) = 0.059\,\Biggl(\frac{F_{\nu}}{\upmu \mathrm{Jy}}\Biggr)\,(1+z)^{-(\alpha+1)}\,\Biggl(\frac{D_L}{\mathrm{Gpc}}\Biggr)^{2}\Biggl(\frac{\nu}{\mathrm{GHz}}\Biggr)^{-\alpha}
,\end{equation}
where $F_{\nu}$ is the flux at the frequency $\nu$, $z$ is the redshift, $D_L$ is the luminosity distance, and $\alpha$ is the spectral index, which we assume to be -0.75. In addition to the sources reported in Table B.1, we collected 67 non-detections from the literature, resulting in upper limits on the SFRs down to $<$0.02\,M$_{\odot}$\,yr$^{-1}$ (GRB~060218, \citealt{Greiner2016}).
\begin{table}[h!]
\centering
\begin{threeparttable}
%\resizebox{cm}{!}{
\centering
\caption[]{Long-duration GRB host galaxies detected in radio so far.}
\begin{tabular}{ccccccc}
\toprule
GRB         &$z$    &$\nu$  &F$_{\nu}$    &L$_{\nu}$    &SFR   &Ref.\\
            &       &(GHz)  &($\upmu$Jy\,beam$^{-1}$)     &(erg\,s$^{-1}$\,Hz$^{-1}$)  &M$_{\odot}$\,yr$^{-1}$   &\\
\midrule
980425      &0.0085 &1.38   &840$\pm$160   &(1.4$\pm$0.3)$\times10^{27}$        &0.08$\pm$0.02    &\cite{Michalowski2012}\\
980703    &0.967        &1.43   &76$\pm$10      &(3.2$\pm$0.4)$\times10^{30}$   &206$\pm$27  &\cite{Berger2001}\\
000418    &1.119        &1.43   &69$\pm$15      &(4.1$\pm$0.9)$\times10^{30}$   &264$\pm$57  &\cite{Berger2003}\\
020819B    &0.41        &3.0    &31$\pm$8       &(1.7$\pm$0.4)$\times10^{29}$   &20$\pm$5  &\cite{Greiner2016}\\
021211      &1.006      &1.43   &330$\pm$31     &(1.5$\pm$0.1)$\times10^{31}$  &982$\pm$82  &\cite{Michalowski2012}\\
031203\tnote{a} &0.105  &1.39   &254$\pm$46     &(7$\pm$1)$\times10^{28}$       &4.5$\pm$0.8  &\cite{Michalowski2012}\\
050223    &0.591        &5.5    &90$\pm$30      &(1.2$\pm$0.4)$\times10^{30}$   &210$\pm$70 &\cite{Stanway2014}\\
051006      &1.059      &3.0    &9$\pm$3   &(5$\pm$2)$\times10^{29}$    &53$\pm$18  &\cite{Perley2015}\\
051022    &0.809        &5.23   &13$\pm$4       &(4$\pm$1)$\times10^{29}$       &61$\pm$19  &\cite{Perley2013}\\
060505   &0.089         &1.38   &76$\pm$35  &1.5$\pm$0.7$\times10^{28}$ &0.9$\pm$0.4    &\cite{Michalowski2015}\\
060729\tnote{b} &0.54   &5.5    &65$\pm$28      &(7$\pm$3)$\times10^{29}$       &123$\pm$53  &\cite{Stanway2014}\\
060814      &1.92       &3.0    &11$\pm$3       &(2.3$\pm$0.6)$\times10^{30}$   &258$\pm$70  &\cite{Perley2015}\\
061121     &1.314       &3.0    &17$\pm$5       &(1.5$\pm$0.4)$\times10^{30}$   &165$\pm$48  &\cite{Perley2015}\\
070306      &1.496      &3.0    &11$\pm$3       &(1.3$\pm$0.4)$\times10^{30}$   &145$\pm$39  &\cite{Perley2015}\\
080207    &2.086        &5.23   &17$\pm$2       &(4.3$\pm$0.5)$\times10^{30}$   &731$\pm$86  &\cite{Perley2013}\\
080517    &0.089        &4.5    &220$\pm$40     &(4.4$\pm$0.8)$\times10^{28}$   &7$\pm$1  &\cite{Stanway2015}\\
090404\tnote{d}     &3.0        &5.23   &11$\pm$3       &(6$\pm$2)$\times10^{30}$       &1074$\pm$293  &\cite{Perley2013}\\
100316D     &0.059      &1.38   &657$\pm$21     &(5.5$\pm$0.2)$\times10^{28}$   &3.5$\pm$0.1  &\cite{Michalowski2015}\\
100621A\tnote{c}    &0.542      &5.5    &120$\pm$32     &(1.3$\pm$0.3)$\times10^{30}$   &229$\pm$61  &\cite{Stanway2014}\\
111005A     &0.013      &1.38   &245$\pm$30     &(9$\pm$1)$\times10^{26}$       &0.06$\pm$0.01  &\cite{Michalowski2015}\\
\midrule
200716C &0.341   &1.4  &1380$\pm$40 &(5.1$\pm$0.2)$\times10^{30}$   &324$\pm$61   &This work\\
\bottomrule
\end{tabular}
\label{tab:lum}
\begin{tablenotes}
\item \textsl{Column 1:} GRB name. \textsl{Column 2:} redshift. \textsl{Column 3:} observing frequency (GHz). \textsl{Column 4:} Flux density ($\upmu$Jy\,beam$^{-1}$) referred to the observing frequency. \textsl{Column 5:} monochromatic luminosity (erg\,s$^{-1}$\,Hz$^{-1}$). \textsl{Column 6:} SFR calculated with the formula provided by \citet{Greiner2016}. \textsl{Column 7:} References. The uncertainties on the monochromatic luminosity and the SFR are derived with the standard formula for the propagation of errors.
\item [a] We used the flux density measurement at 1.39\,GHz from \citet{Michalowski2012}, while the SFR from \citet{Greiner2016} is derived from the flux density at 5.5\,GHz.
\item [b] We used the flux density value from \citet{Stanway2014} at 5.5\,GHz, while \citet{Greiner2016} derived an upper limit for the SFR using the upper limit for the flux density at 1.39\,GHz from \citet{Michalowski2012}.
\item [c] We used the flux density at from \citet{Stanway2014}, while \citet{Greiner2016} used an upper limit for the flux density at 2.1\,GHz.
\item [d] Even though the host galaxy of GRB~090404 was detected by \citet{Perley2013}, the authors stated that an afterglow origin for the observed detection could not be ruled out.
\end{tablenotes}
\end{threeparttable}
\end{table}
\end{appendix}
\end{document}